\documentstyle[12pt]{article}    
\newcommand{\beq}{\begin{equation}}
\newcommand{\eeq}{\end{equation}}
\newcommand{\la}{\langle}
\newcommand{\ra}{\rangle}
\begin{document}

\title{Elastic Chain in a Random Potential: Simulation of the
displacement function $\langle(u(x)-u(0))^2\rangle$ and relaxation.}

\author{Steven Spencer and Henrik Jeldtoft Jensen\\
Department of Mathematics, Imperial College\\
 180 Queen's Gate, London SW7 2BZ\\
United Kingdom}
\maketitle

\begin{abstract}
We simulate the low temperature behaviour of an elastic chain
in a random potential where the displacements $u(x)$ are confined
to the {\it longitudinal} direction ($u(x)$ parallel to $x$)
as in a one dimensional charge density
wave--type problem. We calculate the displacement
correlation function $g(x)=\langle (u(x)-u(0))^2\rangle$ and the
size dependent average square displacement
$W(L)=\langle(u(x)-\overline{u})^2\rangle$.
We find that $g(x)\sim x^{2\eta}$ with $\eta\simeq3/4$
at short distances and $\eta\simeq3/5$ at intermediate distances.
We cannot resolve the asymptotic long distance dependence of $g$
upon $x$.
For the system sizes considered we find
$g(L/2)\propto W\sim L^{2\chi}$ with $\chi\simeq2/3$.
The exponent $\eta\simeq3/5$
is in agreement with the Random Manifold exponent obtained from
replica calculations and the exponent $\chi\simeq2/3$ is consistent
with an exact solution for the chain with {\it transverse} displacements
($u(x)$ perpendicular to $x$).
The distribution of nearest distances between pinning wells and chain-particles
is found to develop forbidden regions.
\end{abstract}

\newpage

\section{Introduction}
An elastic medium in a random static potential models
numerous physical phenomena such as domain walls in impure
magnets, \cite{huse}, flux line lattices
in dirty Type II superconductors in the mixed
state, \cite{larkin1,larkin2,blatter} polymers in a disordered
environment, \cite{polymer} etc. The configuration of the elastic medium
is determined through the competition between the elastic energy of the
medium and the pinning energy due to the inhomogeneities.  This is
a conflict between ordering and disordering. In less than four
dimensions the random potential will win leading to loss of
translational order at zero temperature.\cite{larkin1}

The degree of distortion of the elastic structure grows with the
considered length scale. Two characteristics are often used to
quantify the distortion. One is the displacement correlation function
\begin{equation}
g({\bf r})=\langle(u({\bf r}+{\bf r}_{0})-u({\bf r}_{0}))^{2}\rangle
\label{correlfunc}
\end{equation}
where $u({\bf r})$ denotes the displacement $u$ of a
point in the elastic medium from its ideal lattice position ${\bf r}$
and the angular brackets denote a thermal average,
average over choice of origin, ${\bf r}_{0}$, and an
average over the disorder.

This function is used in collective pinning theory to calculate
the correlated volume, $V_{c}$ \cite{larkin2}, as well as in the theory
of collective creep \cite{feigelman}. Furthermore, it has been used in
attempts to calculate the correlation function of the translational
order of the elastic system \cite{chudnovsky,sudbo}.

There have been many attempts to calculate the displacement function
either by use of scaling arguments (see for instance Ref.
\cite{larkin2,natterman,feigelman})
or using replica techniques \cite{bouchaud1,bouchaud2,giamarchi,korshunov}.
These studies obtain three separate regimes for $g({\bf r})$.
The first regime is the Larkin regime and is defined by
$g({\bf r}) < R_{p}^{2}$, where $R_p$ is the length scale of the
pinning potential. This short distance regime describes
the behaviour within a correlated volume and is
expected to be derivable from perturbation theory \cite{larkin2,feigelman}.
The second regime is often referred to as the Random Manifold
regime, which is defined over the range $R_p^2 < g({\bf r}) < a_0^2$,
where $a_0$ is the ideal lattice spacing. The final regime is
the asymptotic long distance regime, defined by $g({\bf r}) > a_0^2$.
The displacement function scales differently in each of the
three regimes.

There are three important dimensionalities in
the problem. Firstly, there is the actual dimension of the space
that the random potential occupies, $D$. Secondly there is the
dimension of the components of the elastic medium, $d$
(i.e. points are $d=0$, lines $d=1$, planes $d=2$),
and finally there is the
number of dimensions perpendicular to the components of the
elastic medium that the displacements move in, $n$.
For example, $D=2$, $d=n=1$ represents an interface in a two dimensional
random bond Ising model, whilst $D=3$, $d=2$, $n=1$ is a planar interface
in three dimensions, and $D=3$, $d=1$, $n=2$ represents flux
lines in dirty type II superconductors. In general, $D=d+n$. The
precise scaling in each of the three regimes of $g({\bf r})$ will
usually depend on all three of these dimensionalities.

Another measure of the length scale dependence of the distortion is
obtained from the fluctuations in the displacement field averaged
over the whole system and then studied as function of different system
sizes $L$. I.e.,
\begin{equation}
W(L)= \la (u({\bf r}) - \la u\ra)^2\ra
\label{W-of-L}
\end{equation}

An exact solution for $W(L)$ was offered by Huse, Henley, and Fisher
\cite{huse}. This was for an interface in the continuum version
of the two
dimensional random bond Ising
model (i.e. $D=2, d=1, n=1$).
They derived
an exact scaling exponent for the size dependence of
$W(L)$. Namely $W(L)\sim L^{2\chi}$ with $\chi=2/3$.
The discrete version of this model is similar to ours (discussed
below)
in that both consist of a chain of particles but the above
model is subject to {\it transverse} displacements whilst
the displacements in our chain are confined to the {\it longitudinal}
direction.

Here we consider a one dimensional chain of particles
connected by elastic springs
embedded in a static random potential (i.e. $D=1,
d=0, n=1$). Periodic as well as
open boundary conditions are studied. We calculate the displacement
function in Eq. \ref{correlfunc} and the average square displacement in
Eq. \ref{W-of-L} by Monte Carlo annealing.
We find that $g(x)$ exhibits huge fluctuations and that the averaged
quantity is very sensitive to the specific details of the annealing and
averaging procedure used. Hence, one has to be very careful in order to
make sure that thermal equilibrium is achieved and that the results
are reproducible.

Our motivation is to compare our results to the replica calculations
of Bouchaud, M\'ezard and Yedidia (BMY).\cite{bouchaud1,bouchaud2}
For the given dimensionalities, they obtain the following.
In the Larkin regime, they obtain $g(x)\sim x^{2\eta}$ with
$\eta=3/2$.\cite{jpbpc} In the Random Manifold regime, they get the same
scaling form as the Larkin regime but with an exponent
$\eta=3/5$. Finally, in the long distance regime they obtain
$g(x)\sim x^{2\eta}/\sqrt{\log x}$ with $\eta=3/4$.

In the case of periodic boundary conditions we observe two regimes
of different growth of $g(x)$. We can fit our
data to an algebraic form $g(x)\sim x^{2\eta_i}$.
In the short distance regime $\eta_1\simeq3/4$, in
disagreement with the calculation of BMY, and in
the intermediate distance regime $\eta_2\simeq 3/5$ in agreement
with BMY's replica calculation.\cite{bouchaud1,bouchaud2}
We cannot comment on the scaling in the
long distance regime since computational restrictions prevent us
from considering long enough chains.

In contrast to the theoretical expectation we find
that the exponents measured in the simulation do depend on the strength of
the random potential, with $\eta_2$ approaching 1/2 for the stronger
potential. This could possibly be a finite size effect.

It is very difficult to produce satisfactory statistics for the open chain.
However, our simulation
indicates that there is no difference between the open and the periodic
chain. One might anticipate a difference, since in one case the density
of particles in the chain is allowed to change whereas in the other
case the density is fixed. This difference will of course be most
pronounced at long distances when the relative displacement becomes
of order $a_0$. Hence one could argue that our chains are too short to
resolve the difference. Our largest relative displacement is of
order $2a_0$.

We also find that our simulations give $W(L)\sim L^{2\chi}$,
with $\chi\simeq 2/3$. This is consistent with the exact
solution for the transverse chain. One might expect the
two systems to have different exponents, but it is worth
noting that for disorder free cases where the displacements
are due to thermal effects only, the two systems behave with
the same thermal roughening exponent of $\eta=1/2$.
It is also worth noting that super-universal exponents
can occur.
Numerical simulations of Kardar and Zhang \cite{kardar} show that
the exponent $\chi=2/3$ holds for not only the exact solution where
$D=2, d=1, n=1$, but also for $D=3, d=1, n=2$ and $D=4, d=1,
n=3$. Finally, the difference between transverse and longitudinal
displacements will be most important for relative displacements
large compared to the lattice spacing. We can not exclude a cross-over
to a different $\chi$-exponent at much longer distances.

In order to quantify the amount of correlation between the particles
in the elastic chain and the randomly positioned pinning wells we have
studied the distribution of nearest distances, $P(r)$, between particles and
pinning centres. We observe that $P(r)$ develops gaps around values
of the order of the range of the pinning centres. These forbidden regions
vanish with increasing temperature. This behaviour can be understood
in terms of an effective one particle model.

In addition to the spatial behaviour of the system we also investigate the
relaxation times of the system. We have looked for the possible existence of
ergodicity breaking by measuring the relaxation times as function of system
size as usually done in spin glasses \cite{young}. No ergodicity breaking
was observed.

\section{The Model}
Our model consists of a one dimensional chain of $N$ harmonically
coupled particles with positions $x_{i}$. The equilibrium
ideal lattice spacing is $a_{0}$ and the coupling constant
of the elastic interaction is $k$.
The chain is then subjected to a static background
potential created by $N_{p}$ attractive pinning
wells with random positions $x_{i}^{pin}$. The resulting
Hamiltonian for the system is
\begin{equation}
H=\sum_{i=1}^{N} \frac{k}{2}(x_{i+1}-x_{i}-a_{0})^{2} +
\sum_{i=1}^{N} V(x_{i})
\label{hamiltonian}
\end{equation}
We use $a_0$ as our length unit
and $k$ as our energy unit, thus setting $a_0 = k = 1$.
The random potential at a point $x$ is $V(x)$ and it is generated by
\begin{equation}
V(x)=-\sum_{j=1}^{N_{p}}
A_{p}\exp\left(-\frac{(x-x_{j}^{pin})^{2}}{R_{p}^{2}}\right)
\label{epin}
\end{equation}
The pinning wells are represented by attractive Gaussians
of strengths $A_{p}$ and ranges $R_{p}$. We always take the
density of pinning centres as $n_p =N_p /N =1$.

We perform Monte Carlo annealing simulations on this
Hamiltonian in order to study the low temperature properties of
the system.

We now describe the simulation. Standard Metropolis importance sampling is
used \cite{binney}. The initial configuration for the chain is chosen
at random. We then choose a particle at random and
give it a random displacement of $-\gamma \leq dx \leq
\gamma $. The factor $\gamma$ depends on the temperature and is
chosen such that the acceptance rate is kept at about 60\% for all
temperatures. When calculating $g(x)$ we start from a high
initial temperature $T_i$ and anneal the system down to the final
temperature $T_f$. We use an inversely linear time dependence of the
temperature \cite{szu}
\begin{eqnarray}
T(t)=\frac{T_i}{t} & for &t>0.
\end{eqnarray}
The time unit in the above equation is given by $n_{sweep}$ Monte
Carlo sweeps through the lattice. Thus the time unit $\Delta t=1$
corresponds to $Nn_{sweep}$ Monte Carlo steps.
We find that $n_{sweep}=400$ is sufficient since the results
are the same for larger values of $n_{sweep}$.

At the end of the cooling we check that the simulation at $T=T_f$ is in
equilibrium by calculating the specific heat $C=(\langle H^2\rangle
-\langle H\rangle^2)/k_BT^2$ (here $k_B$ is Boltzmann's constant).
At the very low final temperature we
expect $C$ to be given by the equipartition value $C=Nk_B/2$.
We then perform a thermal average of $g(x)$ by sampling the
configuration of the particles every few sweeps through the
lattice.

The results for $g(x)$ is supplemented by the distribution,
 $P(r)$, of distances $r$ between the particles in the chain
and the pinning centres nearest by. This distribution
contains directly information about the degree of correlation
between the particle position and the positions of the pinning
centres.

In addition to the calculation of $g(x)$ we also measure relaxation
times in simulations performed at fixed temperature.
The relaxation times are obtained from the evolution of the
elastic part of the energy, $E_{el}/N$, in Eq. \ref{hamiltonian}.
In these simulations we start out from a random lattice configuration.
We then measure the elastic energy per particle as function of Monte Carlo
time, $E_{el}(t)/N$ \cite{young}.

\section{Results and Discussion}
Firstly, we study a system of length 200
with $A_p=0.01$ and $R_p=0.1$. The displacement function, $g(x)$,
scales differently depending on whether the fluctuations in the
displacement field are due to thermal roughening or disorder
roughening. For a finite temperature, $T_f$, there is a temperature
dependent crossover
length, $l_T$ \cite{thermal}, that separates the two scaling regimes. For
$x<l_T$
thermal effects dominate and $g(x)$ goes linearly in $x$. For
$x>l_T$ disorder induced displacements dominate. Thus for any finite
temperature, disorder will always dominate if long enough length
scales are considered.
For the considered chain, if we consider final temperatures
$T_f \gg A_p$, the pinning potential becomes irrelevant since
the length scale $l_T$ is greater than the system
size and $g(x)$ is given by the exact expression for the case
where there is no disorder in the system and the only
fluctuations are due to temperature:
\begin{equation}
g(x)=k_B T_fx (1 - \frac{x}{N})
\label{pureg(x)}
\end{equation}
where $N=200$ is the length of the chain.
For $T_f \ll A_p$, the pinning potential totally dominates, since
$l_T$ is less than the lattice spacing, and
$g(x)$ deviates from the expression in Eq. \ref{pureg(x)}.
If we then increase the temperature, we can clearly
identify a change in the short distance behaviour towards the
linear behaviour of the pure system.
Unfortunately large fluctuations limit the systems sizes
affordable. For this reason we have not been able to resolve the
temperature dependence of $l_T$. Below we concentrate on the low
temperature region $T\ll A_p$.

We show in Fig. 1 $g(x)$ measured on a periodic system of length
$N=200$ with $A_p=0.01$ and $R_p=0.1$.
The solid curve is obtained by averaging over 600 different
realisations of the random potential using the same annealing
procedure for each realisation. (The dotted line is discussed below).
In order for the Monte Carlo method
to be able to explore as much of the phase space of the Hamiltonian
as possible, we choose an initial temperature $T_f \gg A_p$ so that
the disorder is irrelevant and thermal effects totally dominate
in the initial stages of the
simulation. The initial temperature here is
$T_i=1$. The solid line corresponds to a final temperature of
$T_f=10^{-3}$. We checked that an average over 800 realisations gives the same
curve where as 400 relaisations were significantly different.
A huge number of realisations are needed in order to obtain satisfactory
statistics. This is due to the low dimensionality of the problem.
Fluctuations are very significant as is seen from Fig. 2 where we have plotted
a host of the $g(x)$ curves for a randomly selected subset of
realisations of the random potential. These strong fluctuations
are already present in the pure system (no random background potential).
We studied the pure system in order to check that our annealing procedure
correctly reproduces the exact result in Eq. \ref{pureg(x)} for $A_p=0$.

We also addressed the question of self-averaging. We did this by comparing the
result obtained for $g(x)$ by averaging over many different realisations
of the disorder to the result obtained for one single realisation of
the disorder averaged over the initial configuration. The result
for the average over initial configurations is shown as the dotted line in
Fig. 1. The final temperature is $T_f = 10^{-3}$. The two different
types of average give the same qualitative results. However,
we attribute the slight difference in the curves to a finite size
effect. Since the curve averaged over initial configurations is for
a single realisation of 200 pinning centres, we can expect that averages
of this nature will not sample the full distribution of pinning
configurations as effectively
as the same average for a larger system or for averages over different
realisations of the disorder. Thus, we expect that
the two types of averages will converge in the thermodynamic limit.
I.e., the chain is self-averaging.

In Fig. 3 we show a double logarithmic plot of $g(x)$ for a periodic
chain of length $N=10^3$ with $T_f=10^{-3}$ for different values
of $A_p$ and $R_p$. We also show the result for an open chain.
One can consistently identify two algebraic regimes where we
define $x=x_c$ as the crossover distance and $g(x)\sim x^{2\eta_i}$
in both regimes. We compare these two regimes to the Larkin regime
and the Random Manifold regime of Ref. \cite{bouchaud1}. In
Ref. \cite{bouchaud1} they obtain a third asymptotic regime for
which $g(x)>a_0^2$, where they obtain $g(x)\sim x^{3/2} /
\sqrt{\log x}$. It can be seen from Fig. 3 that the region
$g(x)>a_0^2$ is too small to obtain any
information about whether the predicted scaling in this regime
is correct or not. To study this we would need to consider much
longer chains which we are prevented from doing due to
computational limitations.
The short distance Larkin regime
$x<x_c$ is in all four cases well described by an exponent $\eta_1\simeq0.73$.
This differs from the value exponent $\eta = \frac{3}{2}$ expected by
BMY.\cite{jpbpc}.
The exponent $\eta_2$ in the Random Manifold regime $x>x_c$ is about 0.63
for $A_p = 0.01$. See the solid line ($R_p=0.1$) and the dashed
line ($R_p=0.2$). The arrow marks the value of $x_c$ for
$A_p = 0.01$ and $R_p=0.1$. It should be noted that the crossover distance
$x_c$ always is found to be only a few lattice spacings. The deviation
of this exponent from analytic expectations might be due to the inapplicability
of continiuum elasticity theory at such short length scales.
When $A_p$ is increased to 0.05 the exponent
$\eta_2$ decreases to about 0.57 as is seen from the dot-dashed curve.
These two exponents are very close the exponent $\eta = \frac{3}{5}$
as predicted in Ref. \cite{bouchaud1}.
The open chain appears to behave identically to the periodic chain.
See the dotted curve  which describes an open chain with
$A_p=0.01$ and $R_p=0.1$.
The crossover distance $x_c$ is predicted to be dependent on the
pinning parameters $A_p$ and $R_p$.
The figure illustrates the difficulties one
has in identifying any functional dependence of $x_c (A_p,R_P)$.
It is also assumed \cite{larkin2,feigelman,bouchaud1} that $g(x)$ at the
crossover distance should scale as $g(x_c)\sim R_p^2$. Again,
it is difficult to extract concrete information about
this.

It is interesting to point out that an exponent $\eta_2=1/2$
is obtained perturbatively by considering a continuum Hamiltonian
with a delta-correlated random {\it potential}.\cite{note2} Consider
\begin{equation}
H=\int dx\{{1\over2}\kappa[{du(x)\over dx}]^2+V(x+u(x))\}.
\end{equation}
The extrema condition $\delta H/\delta u =0$ takes the simple form
\begin{equation}
0=-\kappa{d^2 u(x)
\over dx^2}+V'(x+u(x))\approx -\kappa{d^2 u(x)\over dx^2}
+V'(x)
\label{extrema}
\end{equation}
The approximation is correct to lowest order in $V$.
By integration we obtain
\begin{equation}
u(x)= {1\over \kappa}\int_0^x dx' V(x') + u(0).
\end{equation}
The displacement function in Eq. \ref{correlfunc} is then given by
the following pretty expression
\begin{equation}
g(x)={1\over \kappa^2}\int_0^x dx_1\int_0^x dx_2 \la V(x_1)V(x_2)\ra.
\label{perturb}
\end{equation}
Hence, if the random potential is delta correlated $g(x)\sim x^{2\eta}$
with $\eta=1/2$. We measured the correlation function of the background
potential in the configurations generated by the simulation and found
that the correlations decayed to zero over a very small number of springs
in the chain. This decay distance was of about the same value as the
crossover distance of $g(x)$ (see Fig. 1). We were unfortunately not
able to identify the detailed functional form of $\la V(x_1)V(x_2)\ra$.
This argument suggests that the exponent $\eta_1 \approx 0.73$ measured
in the short distance regime is an effect of the short distance
behaviour of the effective correlations of the potential. This perturbative
result might also have some bearing on the behaviour observed in the
intermediate regime. At these distances the random potential can be
considered delta-correlated and if Eq. \ref{perturb} were applicable
one should expect a value for the $\eta$ exponent of $1/2$. This
perturbative value is not very different from the actually measured
$\eta_2\approx 0.57$ to $0.63$. Nor is it very far from the replica
prediction $\eta=3/5$. Perhaps it is appropriate to mention that
perturbation theory is usually only assumed to be applicable
as long as $g(x)<R_p^2$. This is probably correct when the random potential
consists of a very high density ($N_p\gg N$) of very sharp pinning
centres. In the case we have simulated $N_p=N$. Only few particles
are positioned in the steep regions of the pinning wells where a displacement
of order $R_p$ produces significant changes in $V(x+u(x))$.
Hence, the approximation used in Eq. \ref{extrema} might
still be reasonable for most of the particles even when $g(x)$
becomes equal to or somewhat larger than $R_p^2$.

We show in Fig. 4 a double logarithmic plot of $g(L/2)$, together
with $W(L)$ (see Eq. \ref{W-of-L}) for parameters $A_p =0.05, R_p =0.1$
and $n_p =1$. The two solid lines have
both slope $4/3$. One sees that the behaviour
of both quantities are consistent with the exact result for the
{\it transverse} chain $W(L)\sim L^{4/3}$.\cite{huse}
However, one notices also that the data points at $L=10^3$ fall
below the straight line. This might indicate a crossover
to a different behaviour at longer distances. Hence, we cannot
exclude that the asymptotic behaviour of $W$ and $g(L/2)$
for the considered longitudinal displacements might be different
from the exactly solvable transverse case.

The behaviour of $g(x)$ is put in perspective by the results for
the distribution of nearest distances between particles and
pinning centres shown in Fig.5. Either side of the sharp central
peak there is a dip in the distribution showing the existence of
a `forbidden region' were the particles are less likely to sit. These
regions disappear in the high temperature phase (i.e. when
$l_T > N$) were the distribution
is slightly peaked at the centre and decreases monotonically as $r$
increases.
We can account for the forbidden regions by considering the following
simple model. We describe a single particle in the chain and
the pinning centre nearest to it by the following model Hamiltonian
\[
\begin{array}{ll}
H = \frac{k_{eff}}{2}(x-x_{0})^{2}+A_{p}((\frac{x}{R_p})^{2} -1) & |x|<R_{p} \\
H = \frac{k_{eff}}{2} (x-x_{0})^{2} & |x|>R_{p}
\end{array}
\]
Thus we consider the particle to be interacting with a potential well
of depth $A_{p}$, centred at $x=0$ and of range $R_{p}$, and that this
particle is coupled harmonically to the rest of the chain with an
effective coupling constant $k_{eff}$ and an elastic energy zero at
position $x_0$.
The system is restricted in size to $-a \leq (x,x_0) \leq a$, where the
parameter
$a$ is given by the density of pinning centres in the simulations
$2a = \frac{1}{n_p}$.
Since for each particle the parameter $x_0$ will differ, when we
calculate the distribution from the simulations, this is effectively the
same as averaging the above Hamiltonian (or the probability function
$e^{-\beta H}$) over $x_0$.
Thus, we want to calculate
\begin{equation}
\langle P(x)\rangle_{x_0} =\frac{1}{2a}\int^{a}_{-a} \frac{e^{-\beta H(x,x_0)}}
{Z(x_0,\beta)} dx_0
\label{P(x)calc}
\end{equation}
where $Z(x_0,\beta)$ is the usual canonical partition function with
the exception that the $x$ integration is over the finite interval
$-a \leq x \leq a$. As usual $\beta$ is the inverse temperature.
The calculation reduces to a sum of error functions
and the $x_0$ integration has to be done numerically.
The effective spring constant will be $k_{eff}=\alpha k$, where
$1/\alpha$ gives the number of springs which are affected by the displacement
of a single particle \cite{note3,jensen2}.

Fig. 5. shows the distribution $P(r)$
obtained from a simulation with parameters $N=10^3$, $A_{p}=0.01$,
$R_{p}=0.1$, $n_{p}=1$ and $\beta =10^3$. The inset to Fig. 5. shows
the calculated distribution $\langle P(x)\rangle_{x_0}$ with the same
parameters as the simulation and with a renormalisation constant
$\alpha=0.175$. The forbidden regions can be seen clearly in both
cases. It is captivating to notice that the length scale $1/\alpha=5.7$
is identical to the length scale $x_c$ at which $g(x)$  crosses between the
two different algebraic regimes for this set of parameters. See
Fig. 3 where the arrow at the x-axis locates the value $\log_{10}(1/\alpha)$.

It is of interest to measure the relaxation times of the
system in order to decide upon the possibility of ergodicity,
or replica, breaking in the disordered chain \cite{young}.
We measured the time dependence
of the elastic energy $E_{el}(t)$ (given by the first term in
the Hamiltonian in Eq. \ref{hamiltonian}).
The relaxation of $E_{el}(t)$  for different chain lengths is exhibited
in Fig. 6 for $A_p=0.01$, $R_p=0.1$ at a temperature of $T=10^{-3}$.
An average over different realisations of the random potential has
been performed. The temporal behaviour of
$E_{el}(t)$ is independent of the length of the chain. Hence, we
conclude that the disorder is unable to break the ergodicity of the system.

\section{Conclusion}
We have studied a harmonic chain at low temperature subject to a
random potential. Fluctuations, both thermal as well as disorder,
are very significant. This puts unpleasant limits on the system sizes
which one can simulate in a reliable way. In spite of these limitations
we conclude the following concerning the spatial behaviour (at low
temperatures)
of the displacement function $g(x)$. A crossover length $x_c$ exists. For
$x<x_c$ we find $g(x)\sim x^{2\eta_1}$ with $\eta_1\simeq3/4$,
in disagreement with the expected Larkin exponent. For
$x>x_c$ we find $g(x)\sim x^{2\eta_2}$ where $\eta_2\simeq3/5$,
in agreement with replica theory. We are not able to identify the functional
dependence of $x_c$ on the strength and range of the randomly positioned
pinning wells. We do not find any difference in the behaviour of
the open and the periodic chain. We are also unable to consider
chains of sufficient length so to study the appearance of a third
scaling region for $g(x)>a_0^2$.

The average square fluctuations $W(L)\sim L^{2\chi}$ was found to depend
on the length of the chain with an exponent $\chi$ consistent with
the exact result for a {\it transverse} chain $\chi=2/3$. The exponent
might change on length scales beyond our reach.

We studied the distribution $P(r)$ of nearest distance between particles and
pinning wells. As the temperature is lowered (or $A_p$ is increased)
particles become prohibited from a region around $R_p$ away from
the centre of the pinning wells.

The crossover distance $x_c$ and the distribution $P(r)$ can be connected.
In an effective single particle model of $P(r)$ the scale $x_c$ reappears
as the ratio between the effective spring constant and the bare spring
constant.

The system was found to be self-averaging. A study of relaxation times in the
system showed that (as expected in one dimension) no ergodicity breaking occur.

Acknowledgement: We are grateful to J.P. Bouchaud, K.H. Fischer,
P. Holdsworth, and D. O'Kane for stimulating discussions. The work
is supported by the British EPSRC and DRA Malvern.

\newpage
\begin{center}{\bf  Captions} \end{center}
\noindent{ \bf Figure 1.} \\
The displacement correlation function. Solid line corresponds to
an average over 600 realisations of the random potential. The
dotted line is generated by an average over 2300 different initial
starting configurations all annealed down into the same realisation
of the random potential. Parameters are $N=200$, $A_p=0.01$,
$R_p=0.1$, $n_p=1$, $T_i=1$, and $T_f=10^{-3}$.

\noindent{ \bf Figure 2.} \\
A host of the concrete realisations of $g(x)$ entering into the
averaged $g(x)$ shown as the solid line in Fig. 1.

\noindent{ \bf Figure 3.} \\
Double logarithmic plot of $g(x)$ in four different cases.
Solid line represents $A_p=0.01$ and $R_p=0.1$. Dashed line
corresponds to $A_p=0.01$ and $R_p=0.2$. The dot-dashed line belongs
to $A_p=0.05$ and $R_p=0.1$. These three curves are for periodic chains.
The dotted curve  represents the measurement on an open
chain with $A_p=0.01$ and $R_p=0.1$. In all cases $N=10^3$, $T_i=1$,
and $T_f=10^{-3}$. The arrow indicates the location of $\log_{10}(x_c)$
which coincides with the location of $\log_{10}(1/\alpha)$.
See the text for an explanation.

\noindent{ \bf Figure 4.} \\
Double logarithmic plot of the average fluctuations in the displacement
$W(L)$ as function of $L$ (crosses) together with the amplitude
$g(L/2)$ of the displacement correlation function (circles).

\noindent{ \bf Figure 5.} \\
The distribution of nearest distance between particles in the chain
and pinning centres. Parameters are  $A_p=0.01$ for the solid line
and $A_p=0.05$ for the dashed curve. In both cases  $R_p=0.1$, $n_p=1$,
$N=10^3$, $T_i=1$, and $T_f=10^{-3}$.  The insert corresponds to the
solid curve. It is calculated from the single particle model with a
scaling paramenter $\alpha=0.175$.

\noindent{ \bf Figure 6.} \\
The time dependence of the  elastic energy (measured
per particle). The simulation is
run at fixed temperature $T=10^{-3}$. Curves for $N=200$,
$300$, $400$, and $800$ are shown. In all cases $A_p=0.01$, $R_p=0.1$,
$n_p=1$.


\begin{thebibliography}{99}
\bibitem{note3} The idea is the following. Take $n$ springs each of
spring constant $k$. Couple these $n$ springs in a series. The
effective spring constant $k_{eff}$ of the series is $k_{eff}=k/n$.
\end{thebibliography}
\end{document}